\documentclass[final,12p,times,twocolumn
]{elsarticle}

\usepackage{amssymb}
\usepackage{lipsum}

\setcitestyle{square}

\usepackage{amsmath}
\usepackage{amssymb}
\usepackage{graphicx}
\usepackage{amsfonts}
\usepackage{latexsym}
\usepackage{bbold}
\usepackage{wasysym}
\usepackage{calligra}
\usepackage{float}
\usepackage{ulem}
\usepackage{inputenc}
\usepackage{xspace}
\usepackage{url}
\usepackage{epstopdf}
\usepackage{tikz}
\usepackage{amsthm}
\usepackage{soul}
\usepackage{hyperref}

\newcommand{\beq} {\begin{equation}}
\newcommand{\eeq} {\end{equation}}

\journal{Physics Letters B}

\begin{document}
\begin{frontmatter}

\title{Motion of Test Particles in Spacetimes with Torsion and Nonmetricity}


\author[add1]{Damianos Iosifidis}
\ead{damianos.iosifidis@ut.ee}
\author[add2]{Friedrich W. Hehl}
\ead{hehl@thp.uni-koeln.de}

\address[add1]{Laboratory of Theoretical Physics, Institute of Physics, University of Tartu, W. Ostwaldi 1, 50411 Tartu, Estonia.}
\address[add2]{Institute for Theoretical Physics, University of Cologne, 50923 K\"oln, Germany}

\begin{abstract}

We derive the equations of motion of a test particle with intrinsic hypermomentum in spacetimes with both torsion $S$ and nonmetricity $Q$ (along with curvature $R$). Accordingly, $S$ and $Q$ can be measured by  tracing out the trajectory followed by a hypermomentum-charged test particle in such a non-Riemannian background. The test particle is approximated by means of a Dirac $\delta$-function. Thus we find a tangible way to observe and measure the effects of torsion and nonmetricity. Our results are consistent with earlier ones derived by  Obukhov and Puetzfeld (2014) by means of a different method. We apply our insight and evaluate how far-reaching the so-called `geometrical trinity of gravity' really is. 

\end{abstract}

\begin{keyword}
Torsion \sep Nonmetricity \sep Hypermomentum \sep Test Particle \sep Metric-Affine Gravity \sep Non-Riemannian Geometry

\end{keyword}
\end{frontmatter}


\section{Introduction}

The Metric-Affine (gauge theory of) Gravity \cite{MAGPRs} has the remarkable property to relate spacetime torsion $S$ and nonmetricity $Q$ (being the observables) to the microscopic properties of matter, such as the currents of spin, dilation, and shear (the sources). Consequently, in order to be able to observe possible non-Riemannian effects, test matter with internal structure must be considered \cite{Puetzfeld:2007ye,obukhov2014conservation,Neeman:1996zcr}. Torsion and nonmetricity can then be detected by 
deriving  the equations of motion for the momentum and hypermomentum of the particle. Then we can trace out the trajectory of a hypermomentum-charged (i.e. microstructured) test particle in this post-Riemannian arena. Here, by approximating the position of the test particle by a Dirac $\delta$-function, we investigate exactly this problem of finding out the trajectory followed by the microstructured particle.  

Earlier results are due to Puetzfeld and Obukhov (2014). In the general relativistic tradition of Mathisson, Papapetrou, and Dixon, see the review article of Dixon\cite{Dixon:2015vxa}, they extended the methods used in general relativity to metric-affine gravity. They used a multipole expansion  and solved the system up to second order in the corresponding gravitational moments. Their method is exact in principle, but in practice they stopped to make it explicit after the second order, i.e., they eventually could solve the pole/dipole system. Our $\delta$-function approximation may appear less satisfactory. However, our method is rather direct and much less time demanding than the approach of Puetzfeld and Obukhov.

The paper is organized as follows: We first introduce the basic ingredients and set-up our conventions and notation. Then we briefly discuss the conservation laws of Metric-Affine Gravity. Subsequently, in Sec.{\ref{sec}}, by using the semi-classical approximation, we derive the desired path equation. The form the latter is reported also for pure dilation, pure spin, and pure shear charges. 

In Sec.4, we investigate the so-called geometrical `trinity' of gravity as formulated in \cite{BeltranJimenez:2019esp}. We will point out, for the first time, that this `trinity' claim is unjustified for all practical purposes. As soon as one additionally considers the equations of motion, which are an integral part of any general-relativistic type theory of gravity, this `trinity' looses its overall meaning.
  
\section{Conservation laws in metric-affine gravity}

The sources of Metric-Affine Gravity are the canonical and metrical energy momentum tensors along with the hypermomentum tensor. They are defined as
\beq
\Sigma^{\mu}_{\;\;\nu}:=\frac{\partial \mathcal{L}_{M}}{\partial(\nabla_{\mu}\psi^{A})}\nabla_{\nu}\psi^{A}-\delta_{\nu}^{\mu}\mathcal{L}_{M},
\eeq
\beq
t_{\mu\nu}:=-\frac{2}{\sqrt{-g}}\frac{\delta (\sqrt{-g}\mathcal{L}_{M})}{\delta g^{\mu\nu}},
\eeq
and
\beq
\Delta_{\lambda}^{\;\;\mu\nu}:=-\frac{2}{\sqrt{-g}}\frac{\delta (\sqrt{-g}\mathcal{L}_{M})}{\delta \Gamma^{\lambda}_{\;\;\mu\nu}},
\eeq
respectively, where $\mathcal{L}_{M}$ is the matter Lagrangian. The metrical energy-momentum is symmetric by default whereas the canonical one is asymmetric in general.

These energy and hypermomentum tensors excite spacetime curvature, torsion, and nonmetricity which, in our conventions, read, respectively:
\begin{equation}
    R^\lambda{}_{\rho\mu\nu} = 2\partial_{[\mu}\Gamma^\lambda{}_{|\rho|\nu]}+2\Gamma^\lambda{}_{\sigma[\mu}\Gamma^{\sigma}{}_{|\rho|\nu]},
\end{equation}
\begin{equation}
    S_{\mu\nu}{}^\lambda = \Gamma^\lambda{}_{[\mu\nu]},\quad\text{and}\quad  Q_{\lambda\mu\nu}=-\nabla_\lambda g_{\mu\nu}.
\end{equation}

We are working in the exterior calculus framework. Then the invariance of the matter action under diffeomorphisms and the general linear group $GL(4,R)$ leads to the conservation laws \cite{Iosifidis:2020gth,Iosifidis:2021nra,Iosifidis:2021bad}
(here expressed in a holonomic frame):
\begin{gather}
	\frac{1}{\sqrt{-g}}(\nabla_{\mu}-2S_{\mu})(\sqrt{-g}\Sigma^{\mu}_{\;\;\alpha})=\nonumber \\-2 S_{\alpha\mu\nu}\Sigma^{\mu\nu} +\frac{1}{2}R_{\lambda\mu\nu\alpha}  \Delta^{\lambda\mu\nu}-\frac{1}{2}Q_{\alpha\mu\nu}t^{\mu\nu} ,\label{cc2}
 \end{gather}
\beq
	\frac{1}{2 \sqrt{-g}}(\nabla_{\nu}-2S_{\nu})(\sqrt{-g}\Delta_{\lambda}^{\;\;\mu\nu}) =
	\Sigma^{\mu}_{\;\;\lambda}- t^{\mu}_{\;\;\lambda}\label{cc1}.
\eeq
This is the most general form of the conservation laws of Metric-Affine Gravity.

 An alternative form is obtained if we expand out the covariant derivative in terms of the Levi-Civita connection plus the distortion tensor $N^\lambda{}_{\mu\nu}.$ We have
 \beq
 N^\lambda{}_{\mu\nu}:= \Gamma^{\lambda}{}_{\mu\nu}-\widetilde{\Gamma}^{\lambda}{}_{\mu\nu}.
 \eeq
     All Riemannian quantities, that is, the ones computed with respect to the Levi-Civita connection, will be denoted by a tilde. After some elementary algebra, we find:
\begin{gather}\hspace{-28pt}
\widetilde{\nabla}_{\mu}\Sigma^{\mu}_{\;\;\alpha}=(S_{\mu\nu\alpha}-2 S_{\alpha\mu\nu}+Q_{\mu\nu\alpha})\Sigma^{[\mu\nu]}+\frac{1}{2}R_{\lambda\mu\nu\alpha} \Delta^{\lambda\mu\nu}\nonumber\\
+\frac{1}{2}Q_{\alpha\mu\nu}(\Sigma^{(\mu\nu)}-t^{\mu\nu}),
\end{gather}
\beq
{\widetilde{\nabla}}_{\nu}\Delta_{\lambda}^{\;\;\mu\nu}=2(\Sigma^{\mu}_{\;\;\lambda}-t^{\mu}_{\;\;\lambda})-N^{\mu}_{\;\;\alpha\beta}\Delta_{\lambda}^{\;\;\alpha\beta}+N^{\alpha}_{\;\;\lambda\beta}\Delta_{\alpha}^{\;\;\mu\beta} \label{t}.
\eeq
Noting also the relation
\beq
N_{[\nu\mu]\alpha}=S_{\mu\nu\alpha}-2 S_{\alpha[\mu\nu]}+Q_{[\mu\nu]\alpha},
\eeq
Eq.(\ref{cc2}) can be rewritten as
\beq\hspace{-5pt}
\widetilde{\nabla}_{\mu}\Sigma^{\mu}_{\;\;\alpha}=\frac{1}{2} R_{\lambda\mu\nu\alpha}\Delta^{\lambda\mu\nu}+\frac{1}{2}Q_{\alpha\mu\nu}(\Sigma^{(\mu\nu)}-t^{\mu\nu})-N_{\mu\nu\alpha} \Sigma^{[\mu\nu]} .\label{conlaw2}
\eeq
After some trivial algebra, the last equation can be reformulated as
\beq
\widetilde{\nabla}_{\mu}\Sigma^{\mu}_{\;\;\alpha}=\frac{1}{2} R_{\lambda\mu\nu\alpha}\Delta^{\lambda\mu\nu}+N_{\nu\mu\alpha} (\Sigma^{\mu\nu}-t^{\mu\nu}) .\label{SigAlt}
\eeq

The advantage of Eqs.(\ref{t}) and (\ref{SigAlt}) is that they both contain the Levi-Civita covariant derivative, meaning that we can freely move the metric in and out of them and perform the various contractions and raising/lowering of the indices. Moreover, we can fully eliminate $t^{\mu\nu}$ from (\ref{SigAlt}) by employing (\ref{cc1}). Furthermore, we perform a post-Riemannian expansion on the curvature. This results in\footnote{In exterior calculus, the energy-momentum law of Metric-Affine Gravity can be formulated very compactly, see \cite[Eq.(52)]{Neeman:1996zcr}:
\begin{equation}
\widetilde D\,[  \Sigma_\alpha +\Delta^{\beta\gamma}(e_\alpha\rfloor N_{\beta\gamma})]+\Delta^{\beta\gamma}\wedge
(\pounds_{e_\alpha}N_{\beta\gamma})= \tau^{\beta\gamma}\wedge(e_\alpha\rfloor\widetilde R_{\gamma\beta}).\nonumber
\end{equation} Our set of conservation laws here maps to Eqs.($55$) and $(56)$ of \cite{obukhov2014conservation} by the substitutions $\Sigma^{i}_{\;\; k}\mapsto \Sigma_{k}^{\;\; i},\; t_{ij} \mapsto -t_{ij},\;\Delta_{j}^{\;\; ik}\mapsto -2\Delta^{i \;\;\;\; k}_{\;\; j}$, and $\Gamma^{l}_{\;ik}\mapsto \Gamma_{ki}^{\;\;\;l}$.}
\begin{gather}
\widetilde{\nabla}_{\mu}\Sigma^{\mu}_{\;\;\alpha}=\frac{1}{2} \widetilde{R}_{\lambda\mu\nu\alpha}\Delta^{[\lambda\mu]\nu}+\frac{1}{2}\widetilde{\nabla}_{\lambda}(\Delta^{\mu\nu\lambda}N_{\mu\nu\alpha})-\frac{1}{2}\Delta_{\lambda}{}^{\mu\nu}\widetilde{\nabla}_{\alpha}N^{\lambda}{}_{\mu\nu}.\label{Law1} 
\end{gather}
At this stage, our Eqs.(\ref{Law1}) and (\ref{t}) are the main results.

\section{\label{sec}Test particle equations of motion in metric-affine gravity}

An energy-momentum tensor describes the densities of the fluxes of energy and momentum at a certain point of spacetime. For a particle with intrinsic spin, its momentum density ${\mathfrak p}_\nu$ is no longer proportional to its transport velocity $u^\mu$, a fact which can be extracted from the Dirac theory of the electron. The analogous is true more generally for a particle with intrinsic hypermomentum. Hence, using a semiclassical approximation, we arrive for the {\it canonical} energy-momentum tensor at the form\footnote{This ansatz is inspired by the physical interpretation of the canonical energy-momentum tensor: The canonical energy-momentum tensor, according to its physical definition, describes the transport of energy-momentum. If we assume that the momentum density ${\mathfrak p}_{\nu}$ is transported with the flux velocity $u^\mu$, then we have $\Sigma^{\mu}{}_{\nu}=u^{\mu}{\mathfrak p}_{\nu}$. This is a so-called convective ansatz.}
\beq\label{em}
\Sigma^{\mu}{}_{\nu}=u^{\mu}{\mathfrak p}_{\nu}.
\eeq
For the hypermomentum tensor we assume a similar convective form,
\beq\label{hyp}
\Delta_\lambda{}^{\mu\nu}=u^\nu{\mathfrak d}_\lambda{}^\mu,
\eeq
which proves to be a valid ansatz in many actual applications.
In Eq.(\ref{hyp}), ${\mathfrak d}_\lambda{}^\mu$ represents the hypermomentum density.

As  {\it metric} (Hilbert) energy-momentum tensor we take here, by hypothesis, that one of a structureless point particle, see Weinberg \cite{Weinberg:1972kfs} and also Papapetrou \cite{Papapetrou} and Adler, Bazin, and Schiffer \cite{Adler},
\beq
t^{\mu\nu}=\frac{1}{\sqrt{-g}}m u^{\mu}u^{\nu}\frac{d \tau}{dt}\delta^{(3)}(\vec{x}-\vec{x}_{p}(t)).
\eeq
Here $m$ is the mass of the particle and $\vec{x}_{p}(t)$ its position in $3$-space. Analogously, in the limit where the particle becomes pointlike, the above expressions for the canonical energy-momentum and hypermomentum tensors take the forms
\beq
\Sigma^{\mu}{}_{\nu}=\frac{1}{\sqrt{-g}}u^{\mu}p_{\nu}\frac{d \tau}{dt}\delta^{(3)}(\vec{x}-\vec{x}_{p}(t))
\eeq
and
\begin{equation}
\Delta_{\lambda}{}^{\mu\nu}=\frac{1}{\sqrt{-g}} u^{\nu} h_\lambda{}^\mu \frac{d \tau}{dt} \delta^{(3)}(\vec{x}-\vec{x}_{p}(t)), \label{e3}
\end{equation}
 respectively.
We are approximating in this article the test particles as {\it point particles.}  We use the Dirac delta function as technical tool.
Then integrals of densities of some rank-2 tensor field $A^{\mu\nu}$ do make sense since in this case, and only then, the outcome of the integration is a true tensor \cite{Adler}:
\beq
\int  d^{3}x \sqrt{\gamma} A^{\mu\nu}.
\eeq
Here $\gamma$ is the determinant of the spatial metric tensor $\gamma_{ij}$.
For the cases of above, one finds the canonical and the metric energy-momentum tensors as
\beq
\int  d^{3}x \sqrt{\gamma} \Sigma^{\mu}{}_{\nu}=u^{\mu}P_{\nu} \label{e1}
\eeq
and
\beq
\int  d^{3}x \sqrt{\gamma} t^{\mu\nu}=m u^{\mu}u^{\nu} ,\label{e2}
\eeq
respectively. Here
\beq
P_{\nu}:=\int  d^{3}x \sqrt{\gamma} p_{\nu}\delta^{(3)}(\vec{x}-\vec{x}_{p}(t)) \label{e3alt}
\eeq
is the total momentum. Similarly, the total hypermomentum reads:
\begin{equation}
H_{\lambda}{}^\mu :=\int  d^{3}x \sqrt{\gamma} h_\lambda{}^\mu\delta^{(3)}(\vec{x}-\vec{x}_{p}(t)). \label{e4}
\end{equation}

With the definitions of above, we integrate Eq.(\ref{t}) and obtain immediately
\beq\hspace{-6pt}
u^{\mu}P^{\nu}=mu^{\mu} u^{\nu}+\frac{1}{2}\frac{\widetilde{D} H^{\nu\mu}}{d \tau} +\frac{1}{2} u^{\beta}(N^{\alpha\nu}{}{}_{\beta}H_{\alpha}{}^{\mu}-N^{\mu}{}_{\alpha\beta}H^{\nu\alpha}),\label{up}
\eeq
where $\frac{\widetilde{D}}{d \tau}:=u^{\alpha}\widetilde{\nabla}_{\alpha}$. Contracting by $u_{\mu}$, we arrive at the momentum-velocity equation
  \beq\hspace{-6pt}
P^{\nu}=m u^{\nu}-\frac{1}{2}u_{\mu}\frac{\widetilde{D} H^{\nu\mu}}{d \tau}-\frac{1}{2}u^{\mu}u^{\beta}(N^{\alpha\nu}{}{}_{\beta}H_{\alpha\mu}-N_{\mu\alpha\beta}H^{\nu\alpha}) .\label{momve}
  \eeq
 This demonstrates our aforementioned statement that the momentum of the point particle, if hypermomentum is present, is not fully aligned to its velocity! In terms of the full covariant derivative, it assumes a more compact form:
\beq\label{nonaligned}
P_{\lambda}=m u_{\lambda}+\xi_\lambda,\;\;\;\text{with}\;\; \;\xi_\lambda:=-\frac{1}{2}u_{\mu}u^{\nu}\nabla_{\nu}H_{\lambda}{}^{\mu}.
\eeq
 
Now we are going to derive the equation of the path. We simply integrate (\ref{conlaw2}) over a $3$-ball enclosing the point particle, use the semiclassical approximation and equations (\ref{e1})-(\ref{e4}). We assume that curvature and distortion vary slowly in the proximity of the particle. Eventually, we arrive at
 \begin{equation}
\frac{\widetilde{D} P^{\nu}}{d \tau}=\frac{1}{2}H^{\alpha\beta} u^{\gamma}R_{\alpha\beta\gamma}^{\;\;\;\;\;\;\nu}
  -N_{\alpha\beta}^{\;\;\;\;\nu}u^{[\alpha}P^{\beta]}   +\frac{1}{2}Q^{\nu}_{\;\;\alpha\beta}u^{\alpha}(P^{\beta}-m u^{\beta}). \label{path1}
 \end{equation}
 \noindent  We have also used the fact that $\widetilde{\nabla}_{\mu}u^{\mu}=0$ which ensures that there is no accumulation/depletion of particles, given that we are considering a single point particle. Equation (\ref{path1}) yields the trajectory of a test particle that is hypermomentum-charged in spaces with torsion and 
 nonmetricity. Note that our result agrees with the one obtained in \cite{Hehl1971} in the limiting case of vanishing nonmetricity and hypermomentum of a purely spin type, that is, $\Delta_{\alpha\mu\nu}=\sigma_{\alpha\mu}u_{\nu}$, where $\sigma_{\mu\nu}$ is the spin density tensor.  Of course Eq.(\ref{path1}) governs the evolution of momentum along the particle's trajectory. In order to get the actual equation of the path, we must substitute (\ref{momve}) into the latter.
 
 Equation (\ref{nonaligned}) yields
 \beq
\xi^{\nu}=-\frac{1}{2}u_{\mu}\frac{\widetilde{D} H^{\nu\mu}}{d \tau}-\frac{1}{2}u^{\mu}u^{\beta}(N^{\alpha\nu}{}{}_{\beta}H_{\alpha\mu}-N_{\mu\alpha\beta}H^{\nu\alpha}) .
 \eeq
We substitute $P_{\mu}=m u_{\mu}+\xi_{\mu}$ into (\ref{path1}). Then we arrive at the final form of the differential equation whose solution yields the trajectory of the particle:
\begin{gather}\nonumber\hspace{-35pt}
    m \left( \frac{d^{2}x^{\nu}}{d \tau^{2}}+\widetilde{\Gamma}^{\nu}_{\;\;\alpha\beta}u^{\alpha}u^{\beta} \right) =\frac{1}{2}H^{\alpha\beta}u^{\gamma}R_{\alpha\beta\gamma}^{\;\;\;\;\;\;\nu} \\
    -u^{\mu}\widetilde{\nabla}_{\mu} \xi^{\nu} +\Big( \frac{1}{2}Q^{\nu}_{\;\;\alpha\beta}-N_{[\alpha\beta]}^{\;\;\;\;\nu}\Big) u^{\alpha}\xi^{\beta}. \label{path0}
\end{gather}
Here the mass conservation equation $\widetilde{\nabla}_{\mu}(mu^{\mu})=0$ has been employed.\footnote{In general, it is also possible to put Regge trajectories  into Metric-Affine Gravity, where the conservation law for mass does't make sense. Under such conditions, we would have an additive term $-\frac{d m}{d \tau}u^{\nu}$ appearing on the right-hand side of (\ref{path0}).} Noting also the relation $Q_{\nu\alpha\beta}=2N_{(\alpha\beta)\nu}$, the last term simplifies and we arrive at the final expression
{\boldmath\beq\hspace{-5pt}
    m \left( \frac{d^{2}x^{\nu}}{d \tau^{2}}\hspace{-2pt}+\hspace{-2pt}\widetilde{\Gamma}^{\nu}_{\;\;\alpha\beta}u^{\alpha}u^{\beta} \right) =\frac{1}{2}H^{\alpha\beta}u^{\gamma}R_{\alpha\beta\gamma}^{\;\;\;\;\;\;\nu}\hspace{-2pt}-\hspace{-2pt}u^{\mu}\widetilde{\nabla}_{\mu} \xi^{\nu}\hspace{-2pt}+\hspace{-2pt}N_{\beta\alpha}^{\;\;\;\;\nu} u^{\alpha}\xi^{\beta} \label{path}.
\eeq }
The first term on the right-hand side is of the Mathisson-Papapetrou type  \cite{Mathisson:1937zz} (see also \cite{Dixon:2015vxa}) and the rest are additional forces in the form of hypermomentum-derivative and hypermomentum-distortion couplings. Of course for particle with no microsctructure (identically vanishing hypermomentum) this reduces to the usual geodesic equation---as it should. 

To have a complete picture of the dynamics of the test particle we should also have an evolution equation for hypermomentum. But this we already have. Indeed, solving  (\ref{up}) for the derivative of $H^{\nu\mu}$ we derive the evolution equation
\beq\hspace{-9pt}\boxed{
\frac{\widetilde{D} H^{\nu\mu}}{d \tau}=2 u^{\mu}(P^{\nu}-m u^{\nu})+u^{\beta}\Big( N^{\alpha\nu}{}{}_{\beta}H_{\alpha}{}^{\mu}-N^{\mu}{}_{\alpha\beta}H^{\nu\alpha} \Big)} \label{Hdynamics}
\eeq
The latter along with (\ref{path}) forms a coupled system of differential equations that fully describe the dynamics of the test particle. Alternatively, one may use the momentum equation combined with (\ref{Hdynamics}) instead which may also be expressed in the more compact form
\beq\boxed{
\frac{\widetilde{D} P^{\nu}}{d \tau}=\frac{1}{2}H^{\alpha\beta} u^{\gamma}R_{\alpha\beta\gamma}^{\;\;\;\;\;\;\nu}+N_{\beta\alpha}^{\;\;\;\;\nu}u^{\alpha}(P^{\beta}-m u^{\beta}).} \label{pathP}
 \eeq
The system of equations (\ref{pathP}) and (\ref{Hdynamics}) is equivalent to that of ({\ref{path}}) and (\ref{Hdynamics}) and the actual choice is a matter of preference. Let us note an immediate conclusion that can be drawn from (\ref{pathP}). We see that for teleparallel geometries (i.e. $R_{\alpha\beta\gamma}^{\;\;\;\;\;\;\nu}\equiv 0$) there can be no Mathisson-Papapetrou force on the right-hand side of (\ref{pathP}). This observation already puts a limitation on the reach that these theories have. Our result seems to support earlier conclusions, see \cite[p.41]{Hehl:1979fnn}, that within teleparallelism spinning matter is {\it not} allowed for consistency reasons.

Finally, in order to get an even better physical picture, let us perform a post-Riemannian expansion of the curvature tensor that appears on the right-hand side of (\ref{pathP}). A trivial calculation reveals,
\begin{gather}
\frac{\widetilde{D} P^{\nu}}{d \tau}=\frac{1}{2}\sigma^{\alpha\beta} u^{\gamma}\widetilde{R}_{\alpha\beta\gamma}^{\;\;\;\;\;\;\nu}  +\frac{1}{2} H^{\alpha\beta}\frac{\widetilde{D} N_{\alpha\beta}{}{}^{\nu}}{d \tau}-\frac{1}{2}H^{\alpha\beta}u^{\gamma}\widetilde{\nabla}^{\nu}N_{\alpha\beta\gamma}\nonumber \\
+\frac{1}{2}H^{\alpha\beta}u^{\gamma}(N_{\alpha\rho\gamma}N^{\rho\;\;\;\;\nu}_{\;\;\beta}-N_{\alpha\rho}^{\;\;\;\nu}N^{\rho}_{\;\;\beta\gamma})     +N_{\beta\alpha}^{\;\;\;\;\nu}u^{\alpha}(P^{\beta}-m u^{\beta}). \label{xxx}
\end{gather}
or equivalently,
\beq
\frac{\widetilde{D} P^{\nu}}{d \tau}=\frac{1}{2}\sigma^{\alpha\beta} u^{\gamma}\widetilde{R}_{\alpha\beta\gamma}^{\;\;\;\;\;\;\nu}  +\frac{1}{2} \frac{\widetilde{D} (H^{\alpha\beta} N_{\alpha\beta}{}{}^{\nu})}{d \tau}-\frac{1}{2}H^{\alpha\beta}u^{\gamma}\widetilde{\nabla}^{\nu}N_{\alpha\beta\gamma},
\eeq
where $\sigma^{\alpha\beta}:=H^{[\alpha\beta]} $ is the spin part of hypermomentum. Note that only this part couples to the Riemannian curvature $\widetilde{R}_{\alpha\beta\gamma}^{\;\;\;\;\;\;\nu}$. Let us stress out that this is in perfect agreement with the multipolar expansion result of \cite{Puetzfeld:2014qba}.

We shall now become more specific and consider a particle that is charged with dilation, spin, and shear, respectively. 

\subsection{Pure Dilation Case}

Let us consider a point particle that is charged only with dilation. In this case, its hypermomentum is of the form
\beq\label{dil}
\hat{\Delta}_{\lambda}^{\;\;\mu\nu}=\delta_{\lambda}^{\mu}\Delta^{\nu},
\eeq
where $\Delta^{\mu}=\Delta u^{\mu}$ is the dilation current, with $\Delta$ being the dilation charge density. In this case, Eq.(\ref{up}) takes the form
\beq
u^{\mu}P^{\nu}=m u^{\mu}u^{\nu}+\frac{1}{2}g^{\mu\nu}{\rm div}\Delta,
\eeq
where ${\rm div}\Delta:=\frac{1}{\sqrt{-g}}\partial_{\alpha}(\sqrt{-g}\Delta^{\alpha})$ is the divergence of the dilation current. Contracting Eq.(\ref{dil}) by $h_{\mu\nu}=g_{\mu\nu}+u_{\mu}u_{\nu}$, we find as a consistency relation
\beq
{\rm div}\Delta=0. \label{div0}
\eeq
Thus, the dilation current is a conserved vector current.

Moreover, expanding the latter, we have   $0=\widetilde{\nabla}_{\mu}(\Delta u^{\mu})=\dot{\Delta}+\Delta \widetilde{\nabla}_{\mu}u^{\mu}$. Thus, combining Eq.(\ref{div0}) with $\widetilde{\nabla}_{\mu}u^{\mu}=0$, it follows that, quite remarkably, the dilation charge does not change along the particle's motion. Consequently, this $\Delta=const.$ is similar to the constancy of the electric charge. The momentum-velocity equation, in this case, reads 
\beq
P^{\mu}=m u^{\mu}.
\eeq
It is unaffected by the dilation and retains its usual (classical) linear form. It also follows that $\xi^{\nu}=0$. Then the path equation reduces to 
\beq
     \frac{d^{2}x^{\nu}}{d \tau^{2}}+\widetilde{\Gamma}^{\nu}_{\;\;\alpha\beta}u^{\alpha}u^{\beta} =\frac{1}{2 m}\Delta_{\mu}\partial^{[\mu}Q^{\nu]}. \label{pathDilation}
\eeq
Consider now the right hand side of Eq.(\ref{pathDilation}). Since the dilation current is of the form $\Delta^{\mu}=\Delta u^{\mu}$, we see that on the right-hand side of (\ref{pathDilation}) appears a Lorentz-type force, which influences the motion of a dilation charged particle in this non-metric background.

\subsection{Pure Spin Case}

When the particle at hand is only charged with spin, the hypermomentum is antisymmetric in the first two indices, that is, $\hat{\Delta}^{\alpha\mu\nu}=\hat{\Delta}^{[\alpha\mu]\nu}$. In many practical applications it reads as follows:
\beq
\Delta^{[\alpha\mu]\nu}=u^{\nu}\sigma^{\alpha\mu},
\eeq
with $\sigma^{\mu\nu}=-\sigma^{\nu\mu}$ being the spin tensor satisfying the Frenkel condition $\sigma^{\mu\nu}u_{\nu}=0$. Under these circumstances, the momentum-velocity equation (\ref{momve}) takes the form
\beq
P^{\nu}=m u^{\nu}+\frac{1}{2}u_{\mu}\frac{\widetilde{D} \sigma^{\mu\nu}}{d \tau}-\frac{1}{4}\sigma^{\mu\nu}u^{\alpha}u^{\beta}(4 S_{\mu\alpha\beta}+Q_{\mu\alpha\beta}) .\label{pspin}
\eeq
Note that the mass of the point particle is still given by the expression $m=-P_{\mu}u^{\mu}$; this is a consequence of the Frenkel condition and the antisymmetry of the spin tensor. Using these results, the path equation (\ref{path}) now reads
\beq
    m \left( \frac{d^{2}x^{\nu}}{d \tau^{2}}+\widetilde{\Gamma}^{\nu}_{\;\;\alpha\beta}u^{\alpha}u^{\beta} \right) =\frac{1}{2}\sigma^{\alpha\beta}u^{\gamma}R_{[\alpha\beta]\gamma}^{\;\;\;\;\;\;\;\;\nu}+f^{\nu}+N_{\beta\alpha}^{\;\;\;\;\nu} u^{\alpha}\xi^{\beta} , \label{pathSpin}
\eeq
where $\xi^{\nu}$ is the part subsequent to $mu^{\nu}$ in equation (\ref{pspin}) and
\begin{gather}
  f^{\nu}:=-u^{\alpha}\widetilde{\nabla}_{\alpha}\xi^{\nu}=-\frac{1}{2}\frac{\widetilde{D}^{2} \sigma^{\mu\nu}}{d \tau^{2}}  \nonumber \\ +\frac{1}{2}\left( -\frac{\widetilde{D} \sigma_{\mu}{}^{\nu}}{d \tau}+\sigma^{\alpha\nu}u^{\beta}(4 S_{\mu\alpha\beta}+Q_{\mu\alpha\beta})\right) \widetilde{a}^{\mu}
   \\ +\frac{\widetilde{D} \sigma^{\mu\nu}}{d \tau}u^{\alpha} u^{\beta}(4 S_{\mu\alpha\beta}+Q_{\mu\alpha\beta})+\frac{1}{4}\sigma^{\mu\nu}u^{\alpha}u^{\beta}\frac{\widetilde{D}(4 S_{\mu\alpha\beta}+Q_{\mu\alpha\beta})}{d \tau} .\label{f}\nonumber
\end{gather}
Here $\widetilde{a}^{\mu}:=u^{\alpha}\widetilde{\nabla}_{\alpha}u^{\mu}$ denotes the conventional Riemannian acceleration.

\subsection{Pure Shear Case}

If the particle carries only a shear charge, we have a proper traceless hypermomentum tensor \cite{MAGPRs,Blagojevic}, viz.
\beq
\hat{\Delta}^{\alpha\mu\nu}=\hat{\Delta}^{(\alpha\mu)\nu}=J^{\alpha\mu}u^{\nu} \;\;, \;\; \;g_{\alpha\mu}J^{\alpha\mu}=0.
\eeq
Interestingly enough, the above tracelessness condition  ensures that the relation
\beq
m=-P_{\mu}u^{\mu} \label{m}
\eeq
still holds true even in this case. This can be easily seen by contracting (\ref{up}) with $g_{\mu\nu}$. Therefore in all three cases of interest (pure dilation, spin, and shear), the mass relation (\ref{m}) remains valid. In the present shear case, the path equation of the point particle becomes
\begin{gather}
    m \left( \frac{d^{2}x^{\nu}}{d \tau^{2}}+\widetilde{\Gamma}^{\nu}_{\;\;\alpha\beta}u^{\alpha}u^{\beta} \right) =\frac{1}{2}J^{\alpha\beta}u^{\gamma}\Big( g^{\mu\nu}\nabla_{[\gamma}Q_{\mu]\alpha\beta} 
    -2 S_{\gamma}{}^{\nu\lambda}Q_{\lambda\alpha\beta} \Big)\nonumber \\-u^{\mu}\widetilde{\nabla}_{\mu} \xi^{\nu}+N_{\beta\alpha}^{\;\;\;\;\nu} u^{\alpha}\xi^{\beta}. \label{pathShear}
\end{gather}
We further  compute
\beq
\xi^{\nu}=-\frac{1}{2}u_{\mu}\frac{\widetilde{D} J^{\mu\nu}}{d \tau}-\frac{1}{2}u^{\mu}u^{\beta}(N_{\mu\alpha\beta}J^{\nu\alpha}-N^{\alpha\nu}_{\;\;\;\beta}J_{\alpha\mu}).
\eeq
We may then plug back to Eq.(\ref{pathShear}) and arrive at a force with terms similar to (\ref{f}).

To recap, the trajectory of a particle fully charged under hypermomentum (all three currents) obeys equation (\ref{path}), while if it is charged only under dilation, spin, or shear separately, it obeys (\ref{pathDilation}), (\ref{pathSpin}) or (\ref{pathShear}), respectively. The corresponding dynamics of the hypermomentum piece $H^{\nu\mu}$ is contained in eq. (\ref{Hdynamics}).

\section{A geometric 'trinity' of gravity?}

The so-called geometric trinity \cite{BeltranJimenez:2019esp} asserts that gravity can be formulated in three equivalent ways:
\begin{align}
\text{metric theory:}\;\;& R\neq 0,\, S= 0,\, Q=0;    \label{trin1}     \\
\text{metric teleparallelism:}\;\;& R=0, \,S\neq 0,\, Q=0;  \label{trin2}       \\
\text{symmetric teleparallelism:}\;\;& R=0,\, S= 0,\, Q\neq0. \label{trin3}
\end{align}
 A crucial difference among the three cases is that, in the case (\ref{trin1}), hypermomentum vanishes by default, whereas in the two latter cases (\ref{trin2}) and (\ref{trin3}), respectively, hypermomentum will be non-vanishing in general. Thus, for the trinity to be valid, the motion of test particles should be the same in all three cases, otherwise there can be no equivalence. To explore the particle trajectories for each case, it is convenient, for the case (\ref{trin1}), to cast the path equation (\ref{path}) into the form
\beq\label{mot}
  m \left( \frac{d^{2}x^{\nu}}{d \tau^{2}}+\widetilde{\Gamma}^{\nu}_{\;\;\alpha\beta}u^{\alpha}u^{\beta} \right) =\frac{1}{2}\hat{\Delta}^{\alpha\beta\gamma}R_{\alpha\beta\gamma}^{\;\;\;\;\;\;\nu}-g^{\lambda\nu}u^{\alpha} \stackrel{tp}{\nabla}_{\alpha}\xi_{\lambda}.
\eeq
Here $\stackrel{tp}{\nabla}_{\alpha}\xi_{\lambda}=\partial_{\alpha}\xi_{\lambda}- {\stackrel{tp}{\Gamma}}{}\,^{\beta}_{\;\;\lambda\alpha}\xi_{\beta}=\nabla_{\alpha}\xi_{\lambda}-2 S_{\alpha\lambda}^{\;\;\;\;\beta}\xi_{\beta}$ is the covariant derivative with respect to the transposed connection $\stackrel{tp}{\Gamma}{}\,^{\beta}_{\;\;\lambda\alpha}=\Gamma^{\beta}_{\;\;\lambda\alpha}+2 S_{\alpha\lambda}^{\;\;\;\;\beta}=\Gamma^{\beta}_{\;\; \alpha\lambda}$.

Now let us see how far the trinity goes. First of all, we note that for metric theories (\ref{trin1}) the hypermomentum tensor vanishes by default. Then Eq.(\ref{mot}) reduces to the usual geodesic equation:
\beq
 \frac{d^{2}x^{\nu}}{d \tau^{2}}+\widetilde{\Gamma}^{\nu}_{\;\;\alpha\beta}u^{\alpha}u^{\beta} =0.
\eeq
Consequently for the equivalence to exist, the same equation should hold true for metric (\ref{trin2}) and symmetric (\ref{trin3}) teleparallelism as well. Let us investigate. In both cases (\ref{trin2}) and (\ref{trin3}), we have a flat connection, that is, $R^{\alpha}_{\;\;\beta\mu\nu}\hspace{-2pt}\equiv\hspace{-2pt} 0$. Therefore the first term on the right-hand side of Eq.(\ref{mot}) vanishes. The point particle will then follow a geodesic iff  
\beq
u^{\alpha}\stackrel{tp}{\nabla}_{\alpha}\xi_{\lambda}=0. \label{constr}
\eeq

Of course one immediately recognizes that such a condition highly constrains the form of the hypermomentum that these theories allow for. For sensible theories, Eq.(\ref{constr}) must follow from the connection field equations. The instances where this is identically satisfied are the two teleparallel equivalents for which the gravitational actions are just the torsion scalar and the nonmetricity scalar, respectively. Indeed, in both of these cases it can be shown that the connection field equations imply that
\beq
(\nabla_{\nu}-2S_{\nu})(\sqrt{-g}\hat{\Delta}_{\lambda}^{\;\;\mu\nu})=0. \label{dcon}
\eeq
This results in a vanishing $\xi_{\lambda}$ yielding, therefore, the geodesic equation in the end. However, this is a strong constraint on the hypermomentum. Furthermore, by means of Eq.(\ref{cc1}), we see that this condition also demands that the canonical energy-momentum tensor coincides with the metrical one, which cannot be valid in general. It is known, for instance, that for fermionic matter the canonical energy-momentum tensor is not symmetric. Accordingly, it cannot coincide with the metric one. As a result we see that the 'trinity' holds only for highly constrained hypermomentum or for no hypermomentum at all.

Wolf and Read \cite{Wolf:2023rad} even argue that ``The actions of all three theories are equivalent up to a total divergence term—in this sense, all three theories are dynamically equivalent." As we saw from studying the equation of motion of test particles---which are always an integral part of any gravitational theory of a general-relativistic type---this statement is misleading since the equations of motion of test particles destroy this `trinity' radically.

\section{Conclusions}

Using the conservation laws of Metric-Affine Gravity and the semiclassical approximation, we have derived the equations of motion governing the trajectory  of a point test particle, supported by a Dirac $\delta$-function, and carrying hypermomentum in spaces with torsion and nonmetricity. In the most general case (that is, a particle possessing all spin, dilation and shear charges), the system of the differential equations describing the particle's trajectory  is (\ref{path}). The  actual curve swept out by the test particle (influenced by the non-Riemannian features of space) is then found by the solution of the system (\ref{path}). When the particle carries only one of the dilation, spin, and shear currents, the path equation reduces to (\ref{pathDilation}), (\ref{pathSpin}) and (\ref{pathShear}) respectively. 

Finally, we have also addressed the case of the geometrical trinity of gravity and concluded that for the trinity to hold the hypermomentum of the particle must be either vanishing or highly constrained. 

\section{Acknowledgements}

D. I.'s work was supported by the Estonian Research Council grant (SJD14). We are very grateful to Yuri N.~Obukhov (Moscow) and Dirk Puetzfeld (Bremen) for constructive suggestions and for many useful remarks.

\end{document}